\documentclass[12pt]{article}

\usepackage{amsmath,amssymb} 
\usepackage[]{graphicx}
\usepackage{epstopdf}
\usepackage{multirow}
\usepackage{color}
\usepackage{lineno}
\usepackage{ulem}
\usepackage{upgreek}
\usepackage{lipsum}

\usepackage{fullpage}

\usepackage{setspace}
\usepackage{tocloft}
\usepackage[section]{placeins}
\usepackage{capt-of}

\def\blue{\textcolor{blue}}

\def\eps{\varepsilon}
\newcommand{\fraz}{\displaystyle\frac}

\def\##1{{\bf #1}}
\def\=#1{\underline{\underline #1}}

\def\le{\left(}
\def\ri{\right)}

\def\.{\mbox{ \tiny{$^\bullet$} }}
 
\def\ux{\#u_{\rm x}}
\def\uy{\#u_{\rm y}}
\def\uz{\#u_{\rm z}}

\def\eps{\varepsilon}
\def\muo{\mu_{\scriptscriptstyle 0}}
\def\epso{\eps_{\scriptscriptstyle 0}}
\def\lambdao{\lambda_{\scriptscriptstyle 0}}

\def\co{c_{\scriptscriptstyle 0}}

\def\lambdaoBr{\lambda_{\scriptscriptstyle 0}^{\rm Br}}

\def\epsrel{{\=\eps}_{\rm rel}}
\def\epsa{\eps_{\rm a}}
\def\epsb{\eps_{\rm b}}
\def\epsc{\eps_{\rm c}}

\def\Tss{T_{\rm ss}}
\def\Tsp{T_{\rm sp}}
\def\Tps{T_{\rm ps}}
\def\Tpp{T_{\rm pp}}

\def\TLL{T_{\rm LL}}

\def\TRR{T_{\rm RR}}

\def\Ts{T_{\rm s}}
\def\Tp{T_{\rm p}}

\cftpagenumbersoff{figure}
\cftpagenumbersoff{table} 

\begin{document}

\begin{center}
\textbf{Chiral Sculptured Thin Films for Circular Polarization of Mid-Wavelength Infrared Light  }\\

\textit{Vikas Vepachedu and Akhlesh Lakhtakia}\\

{Department of Engineering Science and Mechanics, The Pennsylvania State University, University Park, PA 16802, USA}\\
{akhlesh@psu.edu}

 \end{center}
\begin{abstract}

Being an assembly of identical upright helixes, a chiral sculptured thin film (CSTF) exhibits the circular Bragg phenomenon and can therefore be used as a circular-polarization filter in a spectral regime called the circular Bragg regime. This has been already demonstrated in the near-infrared and short-wavelength infrared  regimes. If two CSTFs are fabricated in identical conditions to differ only in the helical pitch, and if both are made of a material whose bulk refractive index is constant in a wide enough spectral regime, then the center wavelengths of the circular Bragg regimes of the two CSTFs must be in the same ratio as their helical pitches by virtue of the scale invariance of the frequency-domain Maxwell postulates. This theoretical result was confirmed by measuring the linear-transmittance spectrums of two zinc-selenide CSTFs with helical pitches  in the ratio $1$:$7.97$. The center wavelengths were found to be in the ratio $1$:$7.1$, the
deviation from the ratio of helical pitches being explainable
at least in part because the bulk refractive index of zinc selenide decreased a little with wavelength.
We  concluded that CSTFs
can be fabricated to function as circular-polarization filters in the mid-wavelength infrared  regime.

\end{abstract}

 \section{Introduction} 
\label{sec:intro}

The production of circularly polarized light from either an
 unpolarized or a partially polarized source 
involves the removal of either the left-circularly polarized
or the right-circularly polarized  component. One
way of accomplishing this removal is by inserting a rotated linear polarizer  between two
orthogonally oriented quarter-wave plates. Relying on the anisotropy
of the material that it is made of, a quarter-wave plate \cite[pp.~20-21]{Collett} converts linearly
polarized light into circularly polarized light and vice versa. A Fresnel rhomb \cite[p.~49]{Collett} also
converts linearly
polarized light into circularly polarized light, but without reliance on anisotropy.
Instead, its operation is based on the difference between the phase shifts of totally internally
reflected light of two orthogonal linear polarization states.

Another way is to use a slab of a structurally chiral medium (SCM),
exemplified by cholesteric liquid crystals (CLCs) \cite{Chan,deG} and chiral sculptured
thin films (CSTFs) \cite{YK1959}. As an SCM
is 
helicoidally non\-homogeneous along a fixed axis, it displays the circular Bragg phenomenon  
in a specific spectral regime called the circular Bragg regime.
This phenomenon is the almost total reflection of the incident light of one circular polarization state
state but very little reflection of the incident light of the other
circular polarization state, provided that the thickness $L$ of
the SCM is sufficiently high \cite{FLcbp}.  Depending on the 
structural period $P$ as well as the relative permittivity dyadic 
$\epsrel$ of the SCM, the spatiotemporal manifestation of
the circular Bragg phenomenon is the formation of a light
pipe that bleeds energy backward  inside the SCM under
appropriate conditions \cite{GML,GLepjap}.  

The center wavelength $\lambdaoBr$ of the circular Bragg regime depends
on $P$. Typically,  
$\lambdaoBr/P > 1$ for normal incidence   because all eigenvalues of the relative
permittivity dyadics of CLCs and CSTFs exceed unity, and the ratio $L/P$ has to be
in the neighborhood of
$10$ or higher for adequately strong  manifestation of the circular Bragg phenomenon \cite{FLcbp,StJohn}.

Both CLCs \cite{Adams,Scheffer,Isihara,Jacobs} and CSTFs 
\cite{WHL00,PSH08,PMSPL,ZYLZLX} have been demonstrated
to serve as circular-polarization filters in
the visible   and the near-infrared (NIR)  regimes, because
CLCs with $P\lesssim 700$~nm  \cite{Kim2010,Sato,Kulkarni} and  CSTFs with  $P\leq 800$~nm
\cite{PSH08,PMSPL,ELmotl} are commonplace. CLCs with periods as large as about $6000$~nm
have been reported \cite{Li2017}, to our knowledge. But the fabrication of
stable CLCs with the ratio $L/P$
in the neighborhood of $10$ becomes challenging as $P$ increases \cite{Denisov,Kiyoto}.
One reason is that surface alignment forces invoked
by the use of a textured substrate
become less efficacious as the ratio $L/P$ increases \cite{Denisov}. Another reason
is the enhancement of internal stresses which are inimical to the needed structural
chirality of the ordered layering of aciculate molecules becomes poor \cite{Kiyoto}.
CLCs are also sensitive to changes in temperature and pressure \cite{Chan,deG}.
Therefore, although CLCs are suitable for use as circular-polarization filters
also in the short-wavelength infrared (SWIR) regime (wavelength $\lambdao\in(1400,3000)$~nm)
 \cite{Zhang}, similar use in the mid-wavelength infrared (MWIR) regime 
 ($\lambdao\in(3000,8000)$~nm) appears not to be easily possible.

CSTFs are solid-state analogs of chiral liquid crystals. Ideally, a CSTF is
an assembly of parallel helixes of pitch $P$ and rise angle $\chi$.
Structural chirality is thus rigidly built
in the CSTF morphology, so that neither surface alignment forces have to be invoked
nor are internal stresses deleterious. The CSTF morphology is  immune to small 
changes of pressure and temperature, but protection against moisture intake may be necessary \cite{HWM}.

Therefore, with the aim of developing MWIR circular-polarization filters,
we decided to fabricate CSTFs   with $P$ as high as 5880~nm.  
Zinc selenide (ZnSe) was chosen  because its 
bulk refractive index varies quite slowly
with $\lambdao$ in the MWIR regime \cite{Marple} and
because it is easy to deposit CSTFs
\cite{ELmotl,McAtee}
of this material by oblique-angle thermal evaporation \cite{Mattox,STFbook}.
The ratio $L/P$ was kept
as large as practicable
with our thermal evaporation system that was designed for CSTFs to function
in the visible regime and is
definitely suboptimal for industrial use to fabricate very thick CSTFs needed
for the longer wavelengths in the MWIR regime.  We also measured the
normal-incidence
  transmittance characteristics of the fabricated CSTFs.

The plan of this paper is as follows. A brief description of theory underlying the 
  transmission of light in presented in Sec.~2.\ref{reftrans} followed by a discussion
in Sec.~2.\ref{sec:rp} 
of the premise of this research. Fabrication of five different CSTFs is described in Sec.~3.\ref{sec:fab}
and experimental methods for their morphological and optical characterization are
presented in Secs.~3.\ref{sec:mc} and 3.\ref{opt-char}, respectively. Scanning-electron micrographs of all CSTFs fabricated for this research are presented
in Sec.~4.\ref{sec:morph}. Theoretical spectrums of the   transmittances
of a reference CSTF are provided in Sec.~4.\ref{sec:rtr} to understand the very limited
experimental spectrums of the fabricated CSTFs in Sec.~4.\ref{sec:er}. The paper
 ends with some remarks in Sec.~5.

\section{Theoretical Preliminaries}\label{sec:theory}

\subsection{Optical Transmission}\label{reftrans}

Suppose the region $0 \leq z \leq L$ is occupied
by a CSTF, while the half spaces $z \leq 0$ and
$z \geq L$ are vacuous. 
The linear dielectric properties of the CSTF are delineated by the
unidirectionally non\-homogeneous   relative permittivity dyadic \cite{STFbook}
\begin{eqnarray}
\nonumber
&&\=\eps_{\,r} (z,\lambdao) =  \=S_z(h,z,P)\.\=S_y(\chi)\.
\Big[ \epsa(\lambdao) \,\uz\uz +\epsb(\lambdao)\,\ux\ux\\
&&\quad  +\,\epsc(\lambdao)\,\uy\uy\Big]\.\=S_y^{-1}(\chi)\.
\=S_z^{-1}(h,z,P)\, , \,0 \leq z \leq L\, .
\label{epsbasic}
\end{eqnarray}
Here and hereafter,  an $\exp( - i \omega t)$
 dependence on time $t$ is implicit with $\omega=2\pi\co/\lambdao$ as the angular frequency
 and $i=\sqrt{-1}$;
$\co$ is the speed of light in free space;
 and $\ux$, $\uy$, and $\uz$
are the unit vectors in a Cartesian coordinate system.

The helicoidal nonhomogeneity of the CSTF is captured by the
rotation dyadic 
\begin{eqnarray}
\nonumber
&&
\=S_z(h,z,P)=
 \uz\uz + \le \ux\ux+\uy\uy\ri\,\cos\le\frac{2\pi z}{P} \ri
 \\ [5pt]
&&
\qquad +\,h\, \le \uy\ux-\ux\uy\ri\,\sin\le\frac{2\pi z}{P}\ri\,.
\end{eqnarray}
The direction of the non\-homogeneity is parallel to the $z$ axis.  The CSTF
is structurally right-handed when $h = 1$, but  is structurally left-handed when
 $h = -1$. 
The  dyadic
\begin{equation}
\=S_y(\chi) = \uy\uy + (\ux\ux + \uz\uz) \, \cos\chi + (\uz\ux-\ux\uz)\,
\sin\chi
\end{equation}
represents the {\it locally\/} aciculate  
morphology of the CSTF,
with $\chi > 0$~deg being the rise angle.

A 4$\times$4-matrix--based procedure to calculate the linear and
circular remittances of the CSTF of thickness $L$
for  normally incident monochromatic light is explained elsewhere \cite[Chap.~9]{STFbook}
in detail. Using this procedure, we calculated the four linear  
transmittances ($T_{\rm ss,sp,ps,pp}$) and the  
 four circular transmittances ($T_{\rm RR,RL,LR,LL}$) as functions of $\lambdao$. Here,
$T_{\rm sp}$ is the fraction of the incident power transmitted via
an $s$-polarized plane wave when the incident plane wave
is $p$ polarized, $T_{\rm LR}$ is the fraction of the incident power transmitted via
a \textit{L}eft-circularly polarized plane wave when the incident plane wave
is \textit{R}ight-circularly polarized, and so on.

\subsection{Research Premise}\label{sec:rp}
The premise of our research effort can now be explained using the sourceless
versions of the frequency-domain
Maxwell postulates:
\begin{equation}
\left.\begin{array}{l}
\nabla\.\#E(\#r,\lambdao)=0
\\[5pt]
\nabla\.\#H(\#r,\lambdao)=0
\\[5pt]
\displaystyle{\nabla\times\#E(\#r,\lambdao)=i\frac{2\pi\co}{\lambdao}\muo
\#H(\#r,\lambdao)}
\\[5pt]
\displaystyle{\nabla\times\#H(\#r,\lambdao)=-i\frac{2\pi\co}{\lambdao}\epso
\=\eps_{\,r}(z,\lambdao)\.
\#E(\#r,\lambdao)}
\end{array}
\right\}\,.
\label{eq7}
\end{equation}
Here, $\epso$ is the permittivity and $\muo$ is the permeability of
free space.

Let us scale all space isotropically as
\begin{equation}
\#r^\prime = \alpha \#r\,,
\end{equation}
where the scaling parameter $\alpha>1$ is real. If we also scale the free-space wavelength
as
\begin{equation}
\label{lp}
\lambdao^\prime=\alpha\lambdao\,,
\end{equation}
Eqs.~(\ref{eq7}) can be recast as
\begin{equation}
\left.\begin{array}{l}
\nabla^\prime\.\#E(\#r^\prime,\lambdao^\prime)=0
\\[5pt]
\nabla^\prime\.\#H(\#r^\prime,\lambdao^\prime)=0
\\[5pt]
\displaystyle{\nabla^\prime\times\#E(\#r^\prime,\lambdao^\prime)=i\frac{2\pi\co}{\lambdao^\prime}\muo
\#H(\#r^\prime,\lambdao^\prime)}
\\[5pt]
\displaystyle{\nabla^\prime\times\#H(\#r^\prime,\lambdao^\prime)=-i\frac{2\pi\co}{\lambdao^\prime}\epso
\=\eps_{\,r}(z^\prime,\lambdao^\prime)\.
\#E(\#r^\prime,\lambdao^\prime)}
\end{array}
\right\}\,.
\label{eq9}
\end{equation}
Provided that
\begin{equation}
\=\eps_{\,r}(z^\prime,\lambdao^\prime)=
\=\eps_{\,r}(z,\lambdao)\,,
\label{eq11}
\end{equation}
the solutions of Eqs.~(\ref{eq7}) and (\ref{eq9}) shall be identical \cite{Sinclair}.

In other words,  if we can fabricate two CSTFs such
that
\begin{itemize}
\item[(i)] their  helixes have  rise angles
$\chi_1$ and $\chi_2=\chi_1$,
\item[(ii)] their helixes have
pitches    $P_1$ and $P_2=\alpha{P_1}$, 
and
\item[(iii)] their thicknesses are $L_1$ and $L_2=\alpha{L_1}$,
\end{itemize}
then the  transmittances of the first CSTF at 
$\lambdao=\lambda_{0_1}$ shall be the same as  the
  transmittances of the second CSTF at 
 $\lambdao=\lambda_{0_2}=\alpha\lambda_{0_1}>\lambda_{0_1}$ if
 the relative permittivity dyadic $\=\eps_{\,r}^{(1)}$
 of the first CSTF and  the relative permittivity dyadic $\=\eps_{\,r}^{(2)}$
 of the first CSTF satisfy the condition
\begin{equation}
\=\eps_{\,r}^{(2)}(\alpha{z},\lambda_{0_2})=
\=\eps_{\,r}^{(1)}(z,\lambda_{0_1})\,.
\label{eq12}
\end{equation}

Therefore,
if both CSTFs are made by evaporating a  dielectric material whose
bulk relative permittivity is uniform in a spectral regime that encompasses both
$\lambda_{0_1}$ and $\alpha\lambda_{0_1}$, and if  the first CSTF can serve
as a circular-polarization filter in a spectral regime encompassing
$\lambda_{0_1}$, then the second CSTF must serve
as a circular-polarization filter in a spectral regime encompassing
$\alpha\lambda_{0_1}$.  

ZnSe was chosen for evaporation because its 
bulk refractive index $n_{\rm ZnSe}$ varies weakly
with $\lambdao$ in the MWIR regime \cite{Marple,CERDEC}. Since $n_{\rm ZnSe}$
generally decreases as $\lambdao$ increases, as shown in Fig.~\ref{Fig1},
the second CSTF will perform as  a circular-polarization filter in a spectral regime encompassing
wavelengths somewhat smaller than
$\alpha\lambda_{0_1}$.


\begin{center}
    \includegraphics[width=1\columnwidth]{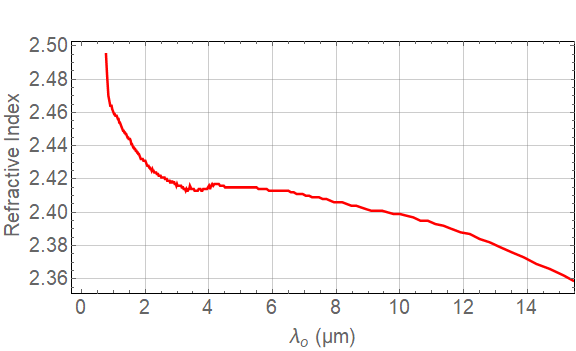}
    \captionof{figure}{Real part of $n_{\rm ZnSe}$ as  a function of $\lambdao$.
    The imaginary part of $n_{\rm ZnSe}$ does not exceed $10^{-6}$ in the same spectral regime  \cite{CERDEC}.  }
   \label{Fig1}
\end{center}


\section{Experimental Methods}

\subsection{Fabrication of CSTFs}\label{sec:fab}

Five different structurally right-handed CSTFs were fabricated using oblique-angle thermal evaporation, the targeted values of  their structural period $P$ and number of periods $L/P$ being listed
 in Table~\ref{tab:chiralDesigns}.

\begin{table}[ht]
\caption{{\bf CSTF Samples Fabricated} }
\label{tab:chiralDesigns}
\begin{center}       
\begin{tabular}{|c|c|c|c|c|c|}  
\hline
\rule[-1ex]{0pt}{3.5ex} Sample & Targeted & Measured   &   &  & Number of  \\
\rule[-1ex]{0pt}{3.5ex}  No. &  $P$ & $P$ &  $L/P$ & $\tau$  & Depositions \\
\rule[-1ex]{0pt}{3.5ex}   &  (nm) & (nm) & & (s)   &   \\
\hline
\rule[-1ex]{0pt}{3.5ex} 1& 366 & 364  & 10 & 5.955 & 3\\
\hline
\rule[-1ex]{0pt}{3.5ex} 2& 732 & 727 & 10 & 12.627 & 5  \\
\hline
\rule[-1ex]{0pt}{3.5ex} 3& 1464 & 1410 & 10 & 25.971 & 10 \\
\hline
\rule[-1ex]{0pt}{3.5ex} 4& 2928 & 2900 & 10 & 52.658 & 20 \\
\hline
\rule[-1ex]{0pt}{3.5ex}  5& 5856 &5880 & 1 & 106.033  & 4  \\
\hline
\end{tabular}
\end{center}
\end{table} 


Thermal evaporation system was carried out in a low-pressure chamber (Torr International, New Windsor, New York). Inside this chamber, the material to be
evaporated is kept in a tungsten boat (S22-.005W, R. D. Mathis, Long Beach, California) which can be heated by passing a current through it. About 15~cm above the boat is a substrate holder 
whose rotations about two mutually orthogonal axes are controlled by two stepper motors. One axis of rotation passes normally through the substrate holder to serve as the $z$ axis, and the second 
serves as the $y$ axis in the substrate ($xy$) plane.
There is also a quartz crystal monitor  (QCM)  in the chamber which has been calibrated to measure the thickness of the film as it is being deposited. Finally, a shutter between the boat and the substrate holder allows the user to abruptly start or halt the deposition as needed. 

99.995 \% pure ZnSe (Alfa Aesar, Ward Hill, Massachusetts) was the material of choice 
for the reasons discussed previously.  The manufacturer supplied ZnSe lumps that were crushed into a fine powder. When crushing the lumps, a respirator mask, gloves, and a lab coat were worn to avoid the toxic effects of ZnSe exposure \cite{MSDS}.

Each CSTF was deposited on two substrates simultaneously, one being either glass or silicon and the other being silicon. The sample
grown on the first substrate was optically characterized.
 If the value of $P$ was chosen for the  circular Bragg regime to lie in the visible regime or the NIR regime or the SWIR regime, a pre-cleaned glass slide (48300-0025, VWR, Radnor, Pennsylvania) was  used as the first substrate. If the value of $P$ was chosen for the  circular Bragg regime to lie in the MWIR regime, silicon was used as the first substrate. 
The morphology of the sample grown on the second substrate was characterized on
a scanning-electron microscope. 

Each substrate was  cleaned
in an ethanol bath using an ultrasonicator for 9~min on each
side; thereafter, the substrate was immediately
dried with pressurized nitrogen gas. Both substrates were secured to the substrate holder using Kapton tape (S-14532, Uline, Pleasant Prairie, Wisconsin), being positioned as close to the center of the holder as possible to ensure that they would be directly above the tungsten boat. The shutter was rotated to prevent any vapor from reaching the two substrates.

To begin the deposition process, the low-pressure chamber was pumped down to approximately 1~$\mu$Torr. Next, the current was gradually increased to $\thicksim$100 A and the shutter was rotated to allow a collimated portion of the ZnSe vapor to reach  the substrates. The deposition rate was manually maintained at $0.4 \pm 0.02$~nm s$^{-1}$, using the QCM. Upon completion of the deposition, the shutter was rotated to prevent the vapor from reaching the substrates
 and the current was quickly brought down to 0 A to prevent any further deposition. After the deposition, the  chamber was allowed to cool for at least 30~min before it was opened. 
 
 Given the desired thickness of the CSTF, the  deposition rate, and the limited amount of ZnSe that
 could be put in the boat, multiple depositions were needed to fabricate the CSTF. The number of depositions for each of the five samples is
shown in   Table~\ref{tab:chiralDesigns}. During each deposition,  a film of maximum thickness
$1500$~nm could be deposited before the ZnSe powder loaded in the boat was depleted. Between these depositions, the boat was refilled with ZnSe powder and, if needed, the quartz crystal in the QCM was replaced. 

During every deposition, the angle of the collimated vapor flux with 
respect to the substrate plane was set at $\chi_v = 20$~deg.
Furthermore, the substrate was rotated about the $z$ axis   in accordance with
the M:1 asymmetric serial bi-deposition technique \cite{McAtee,Pursel} as follows. A subdeposit was made for time $\tau$, followed by a rapid rotation about the $z$ axis by $180$~deg  in   $0.406$~s, followed by another
subdeposit for time $\tau/M$, followed by a rapid  rotation about the $z$ axis by   $183$~deg  in   $0.413$~s.
 This 4-step process was iterated 120 times in order to deposit a single period. Based on a detailed
 comparative study to improve the exhibition of the circular Bragg phenomenon \cite{McAtee}, we
 fixed $M=7$ for all CSTFs. The value of $\tau$ chosen for each CSTF fabricated is  listed in   Table~\ref{tab:chiralDesigns}.

\subsection{Morphological Characterization}\label{sec:mc}

The cross-sectional morphologies of all five CSTFs fabricated
on a silicon substrate were characterized using the FEI Nova\texttrademark~ NanoSEM 630 (FEI, Hillsboro, Oregon) field-emission scanning-electron microscope. Prior to taking the images, the sample was cleaved using the freeze-fracture technique \cite{Severs} so that cross-sectional images could be taken on a clean cleaved edge   free from edge-growth effects. The sample was then sputtered with iridium using a Quorum Emitech  K575X (Quorum Technologies, Ashford, Kent, United Kingdom) sputter coater before imaging.

\subsection{Optical Characterization} \label{opt-char}

Optical characterization of two of the five CSTFs fabricated was performed within 24~h of fabrication, the sample being contained in a desiccator before characterization to prevent degradation due to moisture adsorption. 
A custom-built apparatus was used to measure the linear and circular transmittances 
 of sample~1 \cite{McAtee,Vepachedu},
and  a Vertex 70 spectrophotometer (Bruker, Billerica, Massachussetts)
was used to measure the linear  transmittances of sample~4.

The   transmittance-measurement apparatus for
$\lambdao\in[600\ \text{nm},900\ \text{nm}]$ is described in detail elsewhere \cite{McAtee,Vepachedu}. 
Briefly, light from a halogen source (HL-2000, Ocean Optics, Dunedin, Florida) was passed through a fiber-optic cable and then through a linear polarizer (GT10, ThorLabs, Newton, New Jersey); it was transmitted through the sample to be characterized and then passed through a second linear polarizer (GT10, ThorLabs) and a fiber-optic cable to a CCD spectrometer (HRS-BD1-025, Mightex Systems, Pleasanton, California) \cite{Vepachedu}. The  linear transmittances  were thus measured for normal incidence.
 For measurements of the
circular transmittances, a Fresnel rhomb (LMR1, ThorLabs) was introduced directly after the first linear polarizer and another Fresnel rhomb
directly before the second linear polarizer \cite{McAtee}. All measurements were taken in a dark room to avoid noise from external sources.

The experimental setup for the Vertex 70 spectrophotometer was different. As only one linear polarizer was available, it was used on the incidence side so that only the sums
$\Ts=\Tss+\Tps$ and $\Tp=\Tpp+\Tsp$ were measured for normal incidence.

\section{Results and Discussion}

\subsection{Morphology}\label{sec:morph}

Figure~\ref{Fig2} presents cross-sectional scanning-electron micrographs of  all  five samples listed in Table~\ref{tab:chiralDesigns}. Table~\ref{tab:chiralDesigns} presents the  value of $P$ of every sample fabricated, as estimated from its micrograph. The discrepancy between the targeted and  measured values of $P$ is less than 1\% for four of the five samples, and just 3.7\% for sample~3. The discrepancy is due to the variability inherent in manual control of the deposition rate.
We expect that even better agreement will be obtained after the fabrication process has been optimized for industrial production.


\begin{center}
    \includegraphics[width=1\columnwidth, height=15cm, keepaspectratio]{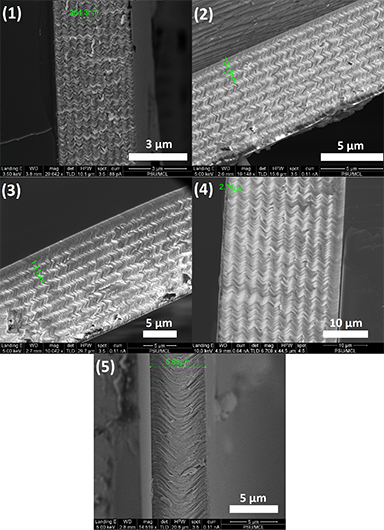}
    \captionof{figure}{Cross-sectional scanning-electron micrographs of all five CSTF samples  listed in  Table~\ref{tab:chiralDesigns}.}
    \label{Fig2}
\end{center}


Each of the five micrographs clearly shows that the fabricated CSTF is an array of very similar helixes.
Even sample~5---which has just one period because the long duration required to fabricate many periods  of that CSTF is infeasible with the resources of an academic laboratory---is an array of helixes. The morphologies of all five samples being the same, except for the scale factor $\alpha$ in Eq.~(\ref{lp}),
we conclude that it is possible to fabricate CSTFs with periods on the order of several micrometers to serve as circular-polarization filters in the MWIR regime.

\subsection{Reference Theoretical  Results\label{sec:rtr}} 

In order to provide theoretical results to serve as a reference
for our experimental findings, we assumed that
the eigenvalues $\epsa$, $\epsb$ and $\epsc$ of $\=\eps_{\,r}$ to have 
single-resonance Lorentzian
dependences \cite{Kittel} on $\lambdao$. Thus,  
\begin{equation}
\label{Lor}
\eps_{\ell}(\lambdao) = 1 + \fraz{p_{\ell}} {  1 + \le 1/N_{\ell}  - i \lambda_{\ell}/\lambdao\ri^{2}  }\, ,
\quad\ell\in\left\{a,b,c\right\}\,,
\end{equation}
with the oscillator strengths  denoted
by $p_{\ell}$. Accordingly, the resonance wavelengths are 
$\lambda_{\ell}  \le 1 + N^{-2}_{\ell}\ri^{-1/2}$ and the linewidths are
$ \lambda_{\ell}/N_{\ell}$,  $\ell\in\left\{a,b,c\right\}$. 
Calculations of all   transmittances, linear as well as
circular, of a reference CSTF for normal incidence were made by setting \cite{ErtenCBP}: $p_{\rm a}=4.7$, $p_{\rm b}=5.2$, $p_{\rm c}=4.6$, $\lambda_{\rm a,c}=260$~nm, $\lambda_{\rm b}=270$~nm, $N_{\rm a,b,c}=130$,  $\chi=50$~deg, $h=1$, $P=324$~nm, and $L=20P$.  

Figure~\ref{Fig3} shows all four circular transmittances of the reference CSTF
for normal incidence. As this CSTF is structurally right-handed, the circular Bragg phenomenon is evident as a trough centered
at $\lambdao\approx 818$~nm  in the
spectrum of $\TRR$, but that feature is absent in the spectrums
of the other three circular transmittances. A comparison of the spectrums of $\TRR$ and $\TLL$ shows clearly
that the reference CSTF can function as a circular-polarization filter at and in the neighborhood
of $\lambdao=818$~nm, which is the center wavelength of the circular Bragg regime.


\begin{center}
       \includegraphics[width=1\columnwidth]{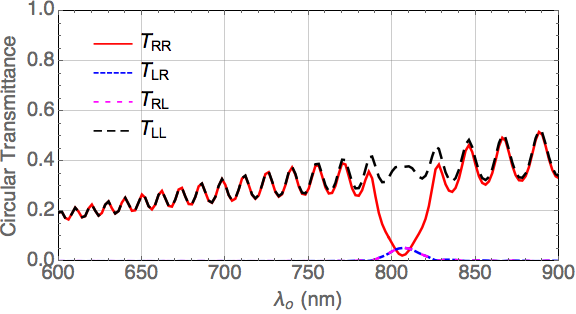}
    \captionof{figure}{Calculated spectrums of the
    circular transmittances of the reference CSTF described in Sec.~4.\ref{sec:rtr}.
 }    \label{Fig3}
\end{center}


The center wavelength is  about $2.52$ times the period  $324$~nm used for the calculations. If the period were increased
by  a factor $\alpha>1$, then the center wavelength would also be increased by the same factor provided that the bulk refractive index of the evaporated material remained invariant, as discussed in Sec.~2.\ref{sec:rp}. ZnSe \textit{almost} fulfills that requirement, as shown in Fig.~\ref{Fig1}. Hence, we can conclude that samples 2--4
should function as circular-polarization filters in progressively longer-wavelength regimes than the one
in which sample~1 does; furthermore, the same would be true if sample~5 were to be fabricated
with a sufficiently large
number of periods \cite{FLcbp,StJohn} instead of just one.

As apparatus to measure the circular   transmittances at longer
wavelengths may not be available, we  provide the spectrums of all four linear transmittances of the reference CSTF in Fig.~\ref{Fig4}.
These spectrums could allow us to properly compare the experimental data on linear transmittances of the larger-$P$ samples with  theory. In Fig.~\ref{Fig4} we notice a distinctive feature in the circular Bragg regime. The troughs of $\Tss$ and $\Tpp$
intersect at  $\lambdao\approx 818$~nm.
If that pattern were to be observed in the experimental results for a given sample in a spectral regime in which the circular Bragg phenomenon is expected,   one could infer the existence of that phenomenon  even if circular transmittances could not be measured.


\begin{center}
       \includegraphics[width=1\columnwidth]{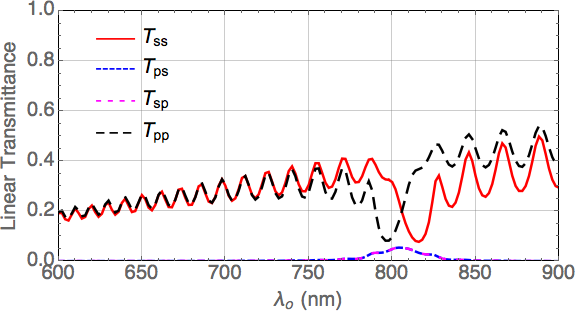}
    \captionof{figure}{Calculated spectrums of the
    linear transmittances of the reference CSTF described in Sec.~4.\ref{sec:rtr}.
  }    \label{Fig4}
\end{center}


\subsection{Experimental  Results}\label{sec:er}

Figure~\ref{Fig5} contains the experimentally determined spectrums of the four
  circular transmittances
of sample~1 ($P=364$~nm, $L=10P$). These spectrums qualitatively match those of the reference CSTF in Fig.~\ref{Fig3}, sample~1 exhibiting a circular Bragg regime centered at $\lambdao\approx780$~nm. At this wavelength, the difference between $\TLL=0.691$ and $\TRR=0.239$ is large enough so that sample~1
can be taken to function as a rejection filter for incident right-circularly polarized light but not for incident left-circularly polarized light. Further improvement may come by increasing the number of periods, i.e., by increasing the ratio $L/P$,
as has been established theoretically \cite{FLcbp,StJohn}.


\begin{center}
    \includegraphics[width=1\columnwidth]{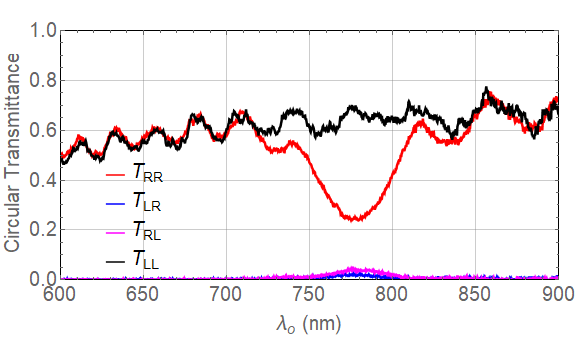}
    \captionof{figure}{Measured spectrums of  the circular
  transmittances  
of sample~1.
    }
    \label{Fig5}
\end{center}


Figure~\ref{Fig6} contains the experimentally determined spectrums of the four linear transmittances
of sample~1. These spectrums qualitatively match those of the reference CSTF in Fig.~\ref{Fig4},
 the troughs of $\Tss$ and $\Tpp$ intersect at $\lambdao=761$~nm in Fig.~\ref{Fig6}. This intersection is very close to 780~nm, the center wavelength of the circular Bragg regime in Fig.~\ref{Fig5} for the same CSTF. Thus, by observing the same intersections for samples with larger $P$,
we can verify the exhibition of the circular Bragg phenomenon by those samples. 


\begin{center}
    \includegraphics[width=1\columnwidth]{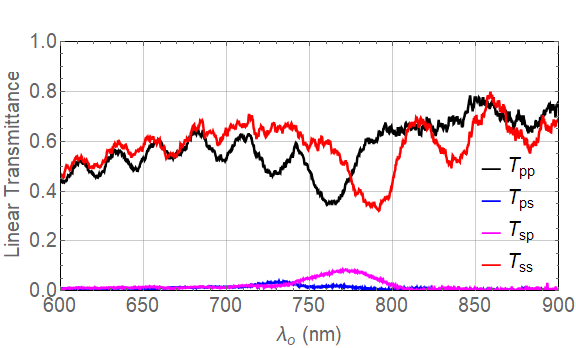}
    \captionof{figure}{Measured spectrums of the linear transmittances  
of sample~1. 
    }
    \label{Fig6}
\end{center}


  Accordingly, we can predict that the center wavelength of the circular Bragg regime of
sample~4 would be $780\alpha=6214$~nm, with $\alpha=2900/364=7.97$. The graphs of $\Ts$ and $\Tp$
in Fig.~\ref{Fig8}
have peaks that intersect at $\lambdao=5512$~nm, which is lower than $6214$~nm. This lowering can be
explained by reduction of
 the bulk relative permittivity of ZnSe  by about 3.05\% in Fig.~\ref{Fig1} as $\lambdao$ increases from
780~nm to 6000~nm.


\begin{center}
    \includegraphics[width=1\columnwidth]{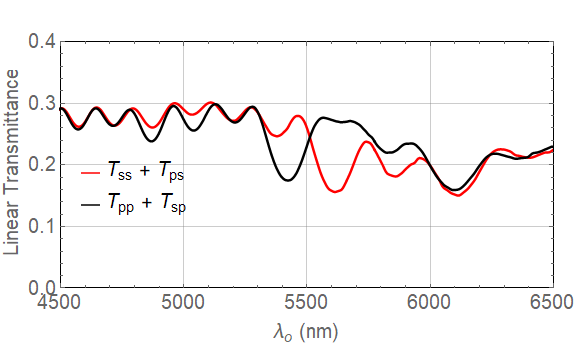}
    \captionof{figure}{Measured spectrums of $\Ts=\Tss+\Tps$ and $\Tp=\Tpp+\Tsp$
      of sample~4.
      }
    \label{Fig8}
\end{center}


Likewise, if a multiple-period version
of sample~5 were to be made, then $\alpha=5880/364=16.15$ and
 the center wavelength of the circular Bragg regime of
sample~5 would be somewhat lower than $780\alpha=12600$~nm.  Thus, we have demonstrated that
CSTFs can be fabricated to serve as circular-polarization filters
in the MWIR regime.

Before closing this section, we must remark  that our apparatus were not appropriate to measure diffuse scattering in nonspecular directions. Nevertheless, our conclusion on the performance of CSTFs as
circular-polarization filters for normal incidence in the MWIR regime holds.
 
\section{Concluding Remarks}
Relying on the scale invariance of the frequency-domain Maxwell postulates \cite{Sinclair}, we selected
a  material whose bulk refractive index is  very weakly dependent on the wavelength.
This material was used as the evaporant material to fabricate  five chiral
sculptured thin films---of structural periods ranging from $\sim$360~nm to $\sim$5900~nm---by
oblique-angle thermal evaporation. The fabrication
conditions of all five CSTFs were thus identical except for a change in scale. Morphological characterization confirmed that
each of the five CSTFs was an assembly of parallel helixes, the helical pitches
in those CTSFs ranging from $\sim$360~nm to $\sim$5900~nm.

As expected from the literature
\cite{FLcbp,ErtenCBP,Kulkarni,Sato}, the measured circular transmittances of the CSTF of the smallest
period demonstrated the circular Bragg phenomenon in the NIR  regime, confirming
that CSTFs of sufficient thickness can function as circular-polarization filters in the same regime.
The linear transmittances of the CSTF of the smallest
period were also measured to identify a spectral feature that would indicate
the occurrence of the circular Bragg phenomenon, this spectral feature also being confirmed
theoretically. The same  feature was experimentally demonstrated to exist in the spectrums of
the linear transmittances of a CSTF whose structural period was about $8$ times larger than that
of the CSTF with the smallest period. Even though the center wavelength of the circular Bragg regime
concomitantly increased by a factor of about $7.1$, at least in part because the bulk refractive index
of the evaporant material decreased a little with $\lambdao$, we can conclude that CSTFs
can be fabricated to function as circular-polarization filters in the MWIR  regime.
\vspace{3mm}

\noindent {\bf Acknowledgments.} VV thanks Pittsburgh Plate and Glass, Inc., for an undergraduate research fellowship. AL thanks the Charles Godfrey Binder Endowment at Penn State for ongoing support of his research
activities.


\end{document}